# Pharmacophore and ligand-guided screening of antibacterial leads targeting antibiotic resistance factor in Gram-negative bacteria

Abi Sofyan Ghifari*[a,b]

[a] *Bioinformatics Research Group, Department of Chemistry, University of Indonesia, Depok 16424, Indonesia;*
[b] *School of Molecular Sciences, The University of Western Australia, Crawley 6009 WA, Australia*

**Abstract:** Pathogenic Gram-negative bacteria have developed resistance to antibiotics due to their ability in creating an envelope on the outer layer of lipooligosaccharides (LOS). The cationic phosphoethanolamine (PEA) decoration of LOS lipid A is regulated by lipid A–PEA transferase (EptA) which may serve as a prominent target for developing new antibiotics. The structural characterization of Neisserial EptA has provided a structural basis to its catalytic mechanisms and ligand recognition that are crucial for inhibitor development. In this study, a combination of pharmacophore– and ligand-based approach has been employed to explore novel potent EptA inhibitors among millions of commercially-available compounds and approved drugs. A total of 8166 hit molecules obtained from ZincPharmer pharmacophore–based screening and PubMed ligand similarity search were further examined through individual two-step semi-flexible docking simulation performed in MOE. Best hits were therefore selected based on their docking score and consensus of the two docking validations. Free energy of binding calculation suggests that the best 20 consensus compounds have a stronger binding affinity than EptA natural substrate PEA. Further interaction analyses of selected eight ligands demonstrate that these ligands have overall more effective interactions with catalytically–essential residues and metal cofactors of EptA. Selected hits can be further analyzed *in vitro* and examined through a pre-clinical trial. This study provides an insight into drug repurposing which may serve as an initial step to develop novel potent EptA inhibitors to combat the virulence of multi-drug resistant Gram-negative bacteria.

## 1. INTRODUCTION

Infections caused by antimicrobial-resistant (AMR) bacteria have become a major global burden in public health from the past several decades and are increasing at an alarming rate [1,2]. Among seven bacterial species that become a global concern as they were substantially developed resistance against multiple antibiotics, five of them are Gram-negative bacteria including Escherichia coli, *Klebsiella pneumoniae*, *Salmonella sp.*, *Shigella sp.*, and *Neisseria sp.* [1]. These bacteria are well known for their involvement in various diseases including gastroenteritis, pneumonia, typhoid fever, shigellosis, as well as gonorrhoea and meningitis, respectively. In this so-called "post-antibiotic era [1]," it is estimated by the middle of 21st century, the mortality rate caused by AMR Gram-negative bacterial infection alone could possibly be increased up to ten million deaths a year [3].

For years, colistin, a peptide antibiotic of polymyxin class, has become the last-resort drug that has been effectively used to treat infections caused by Gram-negative bacteria [4]. Colistin composed of a fatty acid side chain and cyclic polypeptides enriched with cationic amino group that electrostatically interacts with negatively-charged phosphate group of lipopolysaccharides (LPS) layer in the bacterial outer membrane [5]. The interaction eventually leads to the penetration of a lipophilic part of colistin into the outer membrane, ultimately forms pores that cause membrane ruptures and cell lysis [5]. This strategy has been successfully treated infections and sepsis for decades until bacteria that are normally susceptible to colistin have developed a resistance against polymyxins [6]. Colistin resistance has been reported in various Gram-negative bacteria [7–10] and has been characterized to be developed from various mutations of LPS biogenesis and regulation [4,11].

Gram-negative bacteria deploy various strategies for gaining resistance to polymyxin-class antibiotics. Bacteria might enclave polymyxins using anionic polysaccharide capsule [12,13], modify their LPS by adding cationic substances to repel positively-charged polymyxins [14–16], overexpress their outer membrane protein [17], and even completely lost their LPS [11,18]. Among these mechanisms, LPS modification is the most commonly found strategy to achieve polymyxin resistance [6]. LPS can be biochemically modified by attaching positively-charged groups such as phosphoethanolamine (PEA) to counteract colistin, which is the key resistance mechanism in gonococcal and meningococcal pathogens *Neisseria gonorrhoeae* (*Ng*) and *Neisseria meningitidis* (*Nm*) [5,19–22]. In these pathogenic Neisseria, the decoration of lipid A, a lipoglycan moiety that is the major constituent of LPS, with PEA is solely catalyzed by lipid A–PEA transferase A (EptA). EptA (EC 2.7.4.30) belongs to alkaline phosphatase superfamily assists [3]. EptA assists in transferring PEA from phosphatidylethanolamine (PE) to the 1 and 4' phosphate groups of lipid A [5], changing the overall outer negative charges of phosphates to positive charges of PEA which repulse the positively-charged colistin. Not only essential for polymyxin resistance, PEA attachment to LPS has also been associated with the increased meningococcal cell adhesion to human epithelial and endothelial cells [23] and protection of gonococcal Ng lines to immunological systems in mouse genital tract [24]. EptA homologues that also catalyze PEA transfer to lipid A have been characterized in other Gram-negative bacteria [25–28], demonstrating a high conservation of resistance mechanism in this bacterial group. Due to the catalytic activity and direct involvement of EptA in the pathogenicity of Gram-negative bacteria, this enzyme is proposed to be a potential drug target as its inhibition could prevent their resistance against polymyxins [3].

* Address correspondence to this author at the Bioinformatics Research Group, Department of Chemistry, University of Indonesia, Depok 16424, Indonesia.
Email: abi.sofyan@sci.ui.ac.id

Structural characterization of EptA soluble periplasmic domain and its homologues [5,29,30] revealed that the catalytic site contains a zinc ion (Zn2+) which is tetrahedrally coordinated to Glu240, Thr280, Asp452, and His 453 (residue numbers are according to the structure of *Nm*EptA, PDB ID: 4KAV [5]). Thr280, which appeared in the phosphorylated form in *Nm*EptA, is particularly important in EptA catalytic activity as this residue is required for PEA transfer from PE to lipid A via an enzyme–PEA intermediate [3]. A T280A mutagenesis of Mcr-1, an EptA homologue in E. coli, also revealed that this site change increased the vulnerability of bacterial cells to colistin [30], which supports the importance of Thr280 in EptA activity and overall resistance. Therefore, targeting these catalytically active residues of soluble domain of EptA, which responsible of PEA binding and transfer, could be a new way to discover novel antibacterial leads capable of inhibiting PEA decoration of lipid A. Ultimately, the new leads could be further developed and tested to treat infections caused by antimicrobial resistant Neisseria and other Gram-negative bacteria.

Here, I report an early investigation of discovering novel antibacterial leads from databases containing more than a hundred million compounds that target the Gram-negative bacteria resistance factor EptA. This study used a pharmacophore approach which uses structural arrangement and interactions between the target protein and bound molecules or ligands. This approach resulted in more effective inhibitors and stronger predicted interactions with target active site than PEA and some commercially-available antibiotics. The interaction between ligands with the enzyme as well as their binding capacity was further assessed and validated through molecular docking and dynamics studies. As a result, 20 potential leads with a better binding capacity than PEA were obtained from this study. These leads, which are commercially available, may further be tested in vitro and in vivo to develop novel antimicrobial drugs that may treat Gram-negative bacterial–related diseases. The Introduction section should include the background and aims of the research in a comprehensive manner, for the researchers.

## 2. MATERIALS AND METHODS

### 2.1. Protein sequence and structural alignment

Crystal structure of cytoplasmic soluble *Neisseria meningitidis* lipid A–PEA transferase (*Nm*EptA) was selected as a target for this study and obtained from Protein Data Bank (PDB) (http://www.rcsb.org/pdb/), PDB ID: 4KAY [5]. Sequences of Gram-negative bacterial EptA and homologues were obtained from UniProt database (http://www.uniprot.org/) for sequence analysis. Multiple sequence alignment of these protein sequences was done to analyze the conservation of catalytically active residues and was performed in Clustal Omega (https://www.ebi.ac.uk/Tools/msa/clustalo/) [31] using its default parameters. The alignment result was then further analyzed in Geneious software [32] to examine residue consensus and degree of conservation. Structural alignment of three-dimensional Gram-negative bacterial EptA structures was performed to compare structural features of the protein among evolutionarily related bacteria. Structures of EptA periplasmic domain and its homologues were obtained from PDB, namely 4KAY (*Neisseria meningitidis*), 4TN0 (*Campylobacter jejuni*), and 5K4P (*Escherichia coli*). The structures were superimposed and visualized in Maestro [33].

### 2.2. Pharmacophore-based virtual screening and compound library preparation

Potential ligands for were searched based on the EptA structure and pharmacophore approach using

ZincPharmer (http://zincpharmer.csb.pitt.edu/) that utilizes Pharmer algorithm to search potential hits from ZINC database [34]. As the structure of EptA bound with PEA as the substrate is not available, this complex was prepared by re-docking the PEA to EptA active site using Molecular Operating Environment (MOE) [35], which was then uploaded to the server. The compounds used for this virtual screening were deposited in the ZINC database (http://zinc.docking.org/), that contains over 100 million unique purchasable compounds [36]. In total, this screening generated 4865 unique compounds selected from purchasable compounds, natural products, and drugs database of ZINC for further validation. Compound library was also prepared by the ligand-based approach. Using PubChem (https://pubchem.ncbi.nlm.nih.gov/) [37], compounds with structural similarity to PEA were screened and resulted in 407 compounds with at least 80% similarity. In addition, 2000 antibacterials and 894 antibiotics were also included in the library.

All ligands obtained from ZincPharmer and PubChem were downloaded in sdf file format. They were therefore combined into a single database using MOE, generating one MOE database (mdb) file containing a total of 8166 ligands. As the structures of these compounds were still mostly two-dimensional, therefore in a high-energy state, they should be minimized into a lower state of energy. This was performed in MOE using the protocol as described earlier [38]. First, the ligands were protonated to add missing hydrogen in the structure using default wash command. After washing, the partial charges were corrected and the structures were minimized with a 0.05 Å root mean square (RMS) gradient using Merck Molecular Force Field 94 (MMFF94) [39,40]. The library was then ready for docking refinement.

## 2.3. Molecular docking validation and refinement

To start molecular docking protocol, the crystal structure of EptA (PDB ID: 4KAY) was first imported into the MOE system. All water and non-protein molecules, except zinc ion in the catalytic cavity, were removed as the structure underwent energy minimization. The protein structure was minimized using AMBER force field as it was parameterized for macromolecules such as proteins and nucleic acids [41,42]. Missing hydrogens in the structure were initially added using protonate 3D default command and its partial charges were also fixed with AMBER. Then, the structure was minimized with an RMS gradient of no more than 0.1 Å.

For all docking simulations, the catalytic cavity of EptA comprises of tetrad Glu240, Thr280, Asp452, and His 453, as well as the $Zn^{2+}$ (ZN602) cofactor, were designated as docking targets. There are two-step docking protocols that were performed in this study. First, the library of initial 8166 ligands was screened to obtain top 1000 ligands that have the best score. Second, the top 1000 ligands were further refined using duplicate docking with two rescoring steps. The initial docking validation was performed in MOE using Born [43,44] as the solvation method. The placement and refinement methods used in this stage were Triangle Matcher and Forcefield, respectively. This validation retained 30 best ligand positions and removed all duplicates. Top 1000 compounds with the best affinity and binding score, represented by the free energy (ΔG) of binding, were chosen to build a new compound library for further refinement. The refinement was done in duplicates with additional rescoring parameters, namely London dG and Affinity dG, as the first and second rescoring respectively. All refinements retained 100 ligand positions with the highest score and removed any possible duplicates. A total of 20 ligands that appeared in the top 100 of both refinements were then obtained for further interaction analysis.

## 2.4. Interaction analysis and visualization

After molecular docking produced ligand poses with the highest score, the ligand coordinates were saved individually as a MOL file. The ligand coordinate files were then re-opened in MOE and placed in the previously minimized protein coordinate used in the molecular docking protocols. The coordinates of the protein-ligand complex were then saved as a PDB file for interaction analysis and visualization in Maestro. Interaction analyses were done using the Ligand Interaction Diagram function in the Maestro. This tool allows us to observe non-covalent bonds such as hydrogen bond, salt bridge, halogen bond, and aromatic H-bond, pi interactions like pi-cation and pi-pi stacking, as well as to determine the good, bad, and ugly contacts. The ligands were visualized as balls and sticks with carbons in green color, whilst the protein residues represented as thick tube and grey carbons.

## 3. RESULTS AND DISCUSSIONS
## 3.1. Alignment suggests high conservation degree of polymyxin resistance mechanism

Multiple sequence alignment (MSA) of Gram-negative bacterial lipid A–PEA transferase (EptA)

homologues demonstrates a high degree of conservation for the catalytically active tetrad comprises of Glu240, Thr280, Asp452, and His 453. These residues reach a 100% conservation for all 13 assessed EptA homologues (Figure 1a).

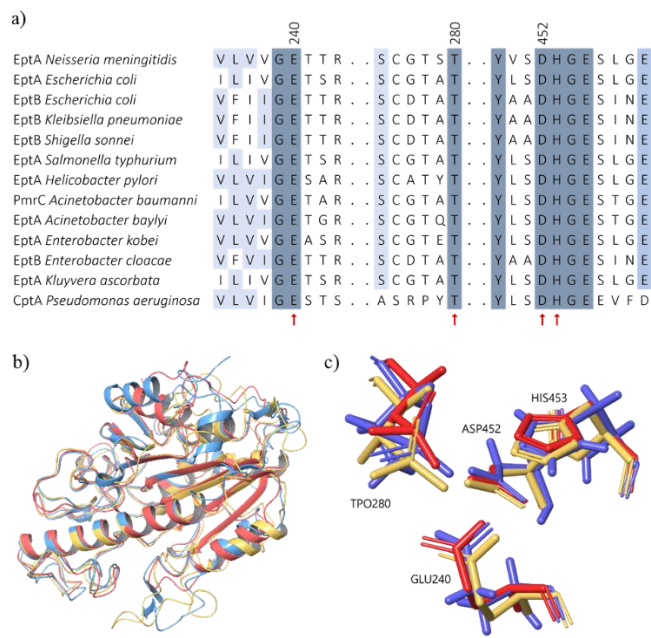

**Fig. (1).** Multiple sequences and structural alignment of EptA homologues in Gram-negative bacteria. a) Sequence alignment of EptA homologues shows high conservation of catalytically active residues (indicated by red arrow) that mediated binding with $Zn^{2+}$ ion and substrate PEA comprise of E240, T280, D452, and H453 (annotation based on *Neisseria meningitidis* EptA [5]). Structural superimposition of EptA homologues available in PDB: *Neisseria meningitidis* EptA (4KAY; red) [5], *Campylobacter jejuni* EptC (4TN0; blue) [29], and *Escherichia coli* EptA/Mcr-1 (5K4P; yellow) [30] demonstrates that: b) protein folds are conserved and c) catalytic tetrads are in the similar position where their side chains face the same direction required for correct binding with PEA.

The catalytic tetrad observed in *Nm*EptA was coordinately-linked with a zinc(II) ion. Although there is no direct evidence that these residues actively bind PEA and promote PEA transfer, a superposition of *Nm*EptA and an alkaline phosphatase, an enzyme that has similar activity to EptA, suggests that the tetrad and the metal ion are needed for substrate binding [5]. This zinc-coordinated tetrad, together with other residues in the catalytic region such as Glu114, His383, and His465, were suggested to assist the binding and transfer of PEA from PE to lipid A [3]. A high conservation of catalytic tetrad in other EptA homologues suggests that these residues are essential for enzymatic activity and the catalytic mechanism is likely to be similarly governed in all Gram-negative EptA analogues.

In addition, structural alignment of three periplasmic EptA homologues available in PDB shows a highly similar architectural arrangement of secondary structure (Figure 1b). The superimposition reveals that all three EptA structures maintain seven-stranded β-pleated sheet that is sandwiched by five α-helices. Further observation on the catalytic tetrad also reveals a similar conservation. The catalytically important Thr280 appeared as a phosphorylated form (abbreviated as TPO) in all structure (Figure 1c), where the phosphate group facing the internal part of the catalytic cavity. Similarly, other three residues Glu240, Asp452, and His453 are also positioned in similar coordinates whilst their side chains also facing the same direction into the inner cavity, which appears to be the substrate binding pocket.

Altogether, these alignments show a significant degree of conservation in sequence and structural aspects and suggest a similar mechanism of substrate binding and catalytic activity of EptA analogues in bacterial resistance to polymyxin antibiotics. Therefore, hit compounds developed from pharmacophore screening and docking that target these residues in this study may be used to treat not only meningococcal diseases but also other Gram-negative bacterial infections.

### 3.2. Ligand library building

Pharmacophore search is a drug discovery approach that employs a spatial arrangement of chemical groups of a ligand within the receptor, usually the target protein. This method allows researchers in exploring structure-activity relationship (SAR) to identify chemical compounds with desired activity [45], such as a strong interaction and inhibitory activity against a protein. Nowadays, this approach is commonly performed computationally and arguably one of the most established and effective methods of virtual screening for rational drug design and discovery [46]. In this study, pharmacophore search was performed in ZincPharmer (http://zincpharmer.csb.pitt.edu/) [34] which accompanies Pharmer pharmacophore search [46] to obtained desired compounds from the largest compound database, ZINC. In contrast to other schemes, Pharmer uses neither fingerprint–based nor alignment-based approaches but rather a completely new indexing approach [46]. Both fingerprint and alignment-based techniques usually assess every structural conformer in a library, therefore require more times as the size of the library increased. On the contrary, the indexing approach used in Pharmer queries conformers based on their geometric complexity, therefore it is independent to the size of the database which significantly reduces the screening time and computational costs [34,46]. Nevertheless, pharmacophore search results need to be further validated in a more accurate technique, i.e. molecular docking to confirm the binding and interaction between ligands and the protein. Therefore, a compound library from virtual pharmacophore screening needs to be established for docking validation.

In this study, the screening of re-docked EptA bound with PEA has provided 4865 screened hits for the compound library. The compounds were obtained from ZINC database [36] and composed of 2000 purchasable compounds, 2000 natural products, and 865 ligands from drug database. In addition to the pharmacophore-searched ligands, the library also included 407 ligand-based compounds that have at least 80% structural similarity to the PEA as well as 2000 antibacterials and 894 antibiotics which were obtained from PubChem [37]. This library then subjected to ligand preparation for docking validation.

As the downloaded structures do not contain hydrogens and are two-dimensional, the compound structure in the library needs to be corrected and energetically minimized. First, the library was protonated to add missing hydrogen atoms in the structure using Protonate 3D command in MOE. Then, the partial charge was corrected and the conformation was energy-minimized, both using the MMFF94 force field [39,40]. MMFF94 is derived from the *ab initio* computational calculation of molecular energies which then parameterized by a wide array of experimentally-determined structures from crystallographic data of receptors and ligands [39]. Therefore, this force field provides a more accurate approximation of structural conformation, that has a low energy state, in an actual wet experimental condition. The compound library then prepared for an initial docking validation.

### 3.3. Docking confirms inhibitory capacity of screened ligands

After the ligand library has been prepared and energy minimized, it underwent first validation of molecular docking. The library of a total 8166 ligands subjected to the active site of EptA soluble domain (4KAY). Initially, the protein structure was prepared similarly to the ligand preparation. All water and non–protein molecules, excluding the $Zn^{2+}$ ion adjacent to TPO280 (ZN1), were removed from the structure.

Missing hydrogens were then added to the structure, while partial charges and energy state were corrected using AMBER99 force field [47]. AMBER99 is a force field built from restrained electrostatic potential (RESP) approach [47], which is parameterized for organic and large biological molecules [41,42,47]. In the structure minimized with a force field, all atoms will find new positions that have the lowest energy. However, the minimized structure should have no more than 0.1 Å of RMS gradient as the higher gradient may possibly deviate from the crystallographic structure and no longer representative for rational drug discovery. In this minimization, the Born approach was used as a solvation model, which generates lower energy state than other available solvation methods such as distance-based and gas-phase.

After the protein attained its lowest possible energy, the docking was then set up. The docking was done in a semi-flexible manner, in which the protein structure remains rigid whilst the ligands can flexibly find a suitable position to obtain the best interaction with target receptors. The placement of ligands into the receptors used Triangle Matcher algorithm which places ligands based on their molecular group charges and spatial arrangement. The poses between atoms in the ligand and the receptor area of the target protein were illustrated as a triangle, where the interaction strength and fitness of these triangles are calculated [48]. In this first docking validation, 30 best ligand positions were retained without duplicates to determine interaction score of ligands with the protein. From this result, top 1000 ligands with lowest Gibbs free binding energy were then extracted to a new ligand database file for further docking refinement.

Docking refinement was performed to further assess the ligand capability to bind the active site and select the ligands with highest inhibitor profile. The refinement was done in duplicates with Triangle Matcher and forcefield were selected as the methods of placement and refinement, respectively. Both refinements used two-step rescoring, namely London dG and affinity dG to calculate the free Gibbs energy of binding ($\Delta G_{binding}$) released from the docking of ligand to the target receptor. By contrast with the initial docking validation which retained 30 best positions, in these refinements 100 best ligand positions were retained to gain more ligand conformations that create strong interaction with the protein which in turn improve the scoring accuracy. From two produced docking databases, 20 ligands that consensually appeared in the top 100 of both libraries were selected (Table 1) for subsequent molecular interaction analyses.

According to binding energy calculation, all 20 ligands significantly have a better binding affinity, ranging from one-half to two times more spontaneous than the EptA natural substrate PEA. Ligand origins are also equally diverse, demonstrating that both pharmacophore and ligand-based approaches produce ligands with a good inhibition profile. The structures of compounds screened from ZincPharmer pharmacophore screening were obtained from ZINC database [36], whilst compounds selected from PEA similarity search were attained from PubChem [37].

**Table (1)**. Free energy binding score of the best 20 consensus ligands obtained from molecular docking refinement protocols

| No | ID | ΔG1 (kcal/mol) | ΔG2 (kcal/mol) | Average ΔG ± std. | Screening |
|---|---|---|---|---|---|
| 1 | 135135840 | -6.935 | -6.935 | -6.935 | Pubchem |
| 2 | 7849111 | -6.450 | -6.238 | -6.344 ± 0.150 | Pubchem |
| 3 | ZINC35000840 | -6.145 | -4.585 | -5.365 ± 1.103 | ZincPharmer |
| 4 | 135136215 | -6.136 | -5.288 | -5.712 ± 0.600 | Pubchem |
| 5 | 26715434 | -6.082 | -5.839 | -5.960 ± 0.172 | Pubchem |
| 6 | 127259617 | -5.873 | -4.918 | -5.395 ± 0.675 | Pubchem |
| 7 | ZINC08101116 | -5.796 | -5.115 | -5.456 ± 0.481 | ZincPharmer |
| 8 | ZINC35000842 | -5.585 | -5.008 | -5.296 ± 0.408 | ZincPharmer |
| 9 | ZINC08101115 | -5.580 | -5.520 | -5.550 ± 0.042 | ZincPharmer |
| 10 | ZINC08101114 | -5.563 | -4.830 | -5.196 ± 0.519 | ZincPharmer |
| 11 | 49661784 | -5.531 | -5.001 | -5.266 ± 0.375 | Pubchem |
| 12 | ZINC13516814 | -5.458 | -5.317 | -5.387 ± 0.099 | ZincPharmer |
| 13 | ZINC08101117 | -5.357 | -4.779 | -5.068 ± 0.409 | ZincPharmer |
| 14 | ZINC89490588 | -5.337 | -4.847 | -5.092 ± 0.346 | ZincPharmer |
| 15 | ZINC08214590 | -5.212 | -4.902 | -5.057 ± 0.219 | ZincPharmer |
| 16 | 171571760 | -5.135 | -4.977 | -5.056 ± 0.111 | ZincPharmer |
| 17 | 7848549 | -5.021 | -4.538 | -4.779 ± 0.342 | Pubchem |
| 18 | 134977160 | -5.021 | -4.757 | -4.889 ± 0.187 | Pubchem |
| 19 | ZINC08215878 | -5.020 | -4.625 | -4.823 ± 0.279 | ZincPharmer |
| 20 | 51091571 | -5.000 | -5.187 | -5.093 ± 0.133 | Pubchem |
| PEA | Phosphoethanolamine | -3.237 | -2.790 | -3.013 ± 0.316 | N/A |

### 3.4. Interaction analyses and visualization of receptor-ligand binding

Currently, there are no crystallographic EptA structures containing in-bound PEA deposited to PDB. Therefore, to hypothetically assess the interaction between the substrate PEA and EptA, PEA was re-docked to the whole surface of the EptA soluble domain structure (4KAY) using MOE to evaluate its binding activities with EptA binding residues. Based on this docking, the pose that has the lowest free energy of binding was produced when the phosphate group of PEA facing inward the known EptA binding pocket. EptA PEA binding site is an open and shallow pocket enriched in polar residues with the catalytic tetrad Glu240, Thr280, Asp452, and His453 as well as coordinately linked $Zn^{2+}$ cofactor positioned at the bottom of the pocket. As the most hydrophilic groups in PEA, the phosphate group apparently capable to strongly interact with the EptA binding pocket and active residues by forming electrostatic bridges with polar residues and the zinc ion (Figure 2).

All three negatively-charged oxygens of the phosphate form electrostatic contacts with the $Zn^{2+}$ cofactor (ZN2), which is suggested to be essential for EptA catalytic activity in transferring PEA to lipid A and stabilizing phosphatized Thr280 [5]. Crystallized EptA structure used in this study (4KAY) contains three zinc ions inside the protein [5]. While one of the zinc ions (ZN3) is situated at around 5.5 Å from the catalytic tetrad, the other two are located within the catalytic site and make contacts with catalytically active residues [5]. The first $Zn^{2+}$ (ZN1) was observed to interact with Asp452 and was presumed to be an alternative active site of the second zinc ion (ZN2), which is the main catalytic site of EptA and located at around 4.4 Å from the ZN1 [5].

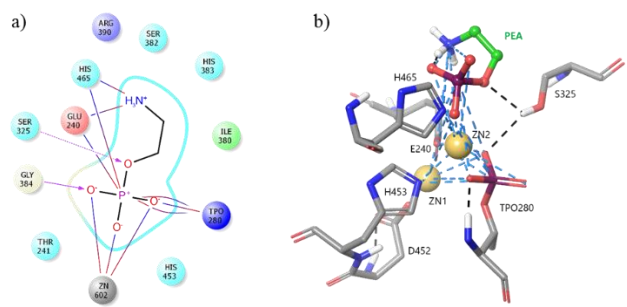

**Figure (2)**. a) two-dimensional and b) three-dimensional interaction analyses of PEA (shown in ball and stick representation and green carbon) and EptA binding site (grey carbon). Salt bridges and other electrostatic interactions are represented as blue dashed lines, whereas hydrogen bonds are shown in black.

Complementary salt bridges are also formed by the phosphate with catalytic residues TPO280 and Glu240. The partially positive phosphor electrostatically interacts with the carbonyl oxygen of Glu240 side chain, whilst a negatively charged oxygen complements the phosphor atom of phosphatized Thr280. In addition, hydrogen bonds are also formed between the phosphate group with two residues of EptA binding surface. The polar hydrogen of Ser325 hydroxyl group forms a hydrogen bond with the carbon-bound phosphoester oxygen, whereas the hydrogen of Gly384 amino backbone forms bond with one of the negatively charged oxygens of the PEA phosphate group. This complex which contains desirable interactions between PEA and EptA active site and therefore used as a template for pharmacophore screening to search compounds with comparable interactions to PEA but a stronger overall binding activity.

After the best 20 consensus ligands with a better docking score than PEA have been determined, their molecular interactions with the receptor were identified using MOE and Maestro. Overall, most hits share similar interaction mode with PEA, in which the polar group(s) is situated inside the EptA binding pocket, whereas the hydrophobic part of the compound fills the binding hallow and make hydrophobic contacts with the pocket wall moieties.

### 3.4.1. Interaction analyses of ligand-based compounds

Ligand 4, 11, 17, and 20 are all obtained from PEA similarity search in PubChem [37]. These compounds have similarities in structures as well as their interaction with the receptor. These ligands have at least four carbon rings with polar groups like hydroxyl group that are attached to these rings. These properties make these compounds capable of binding the enzyme's polar residues and zinc ions as well as making contacts with hydrophobic residues surrounding the catalytic cavity wall. Further structural search against PubChem database revealed that these four compounds are approved and registered antibacterials which belong to the same family of drugs: tetracyclines, more specifically classified as pro-tetracycline drugs [49].

Tetracycline drugs share molecular features, which then determine their minimum pharmacophore [50]. Tetracyclines comprise of a fused tetracyclic core [50], hence the name, which decorated with various active groups serve to enhance their antibacterial activities. Tetracyclines act as an antibacterial by disrupting protein translation process, specifically in binding 30S ribosomes. This prevents aminoacyl transfer RNA to bind the messenger RNA and ribosome complex [49]. In general, tetracyclines show activity against a broad spectrum of bacteria as well as to some fungi and protozoan genera [50], even though they are more commonly used to combat many Gram-negative bacterial infections, including *Escherichia coli*, *Enterobacter*, *Klebsiella*, and *Neisseria* [49]. Although tetracyclines activities against bacterial protein synthesis process are well established, their actions against bacterial antibiotic resistance mechanisms are yet to be examined.

Molecular interaction analyses of these four ligands revealed that all ligands interact with at least two catalytically important residues of EptA. Ligand4, which is better known as meglucycline, interacts with all catalytic tetrad as well as two catalytically essential zinc ions (Figure 3a). At the bottom of the active site, two hydroxyl groups make electrostatic contact, each with ZN1 and ZN2. The polar hydrogen atom of one of these hydroxyl groups also forms a hydrogen bond with an oxygen atom of TPO280 phosphate group. A salt bridge is formed between positively-charged nitrogen and the carboxylate oxygen of the catalytic Glu240.

In its best pose, the tail of Ligand4 is bent towards the catalytic site which allows this ligand to establish additional hydrogen bonds with the catalytic Asp452 and His453. A hydrogen of the amine linker creates a hydrogen bond with the carbonyl oxygen of Asp452 side chain, whilst the hydroxyl group at the end of the molecule makes a hydrogen bond with the primary chain oxygen of His453. Other hydrogen bonds are also formed, mostly with other polar residues, including Thr241, Thr242, Ser451, His378, Thr379, and Tyr449. As a result, even though Ligand4 forms fewer hydrophobic contacts, Ligand4 has the most negative binding free energy (-5.712 kcal/mol) compared to other tetracycline ligands due to its strong interactions with all catalytically important residues and cofactors as well as hydrogen bonds with other polar residues of EptA.

Ligand11 has an average of -5.266 kcal/mol binding free energy with EptA and is registered as lymecycline in most drug and compound databases. Similar to Ligand4, Ligand11 also establishes interactions with most of the catalytically pivotal residues and ions, except for Glu240 and ZN1 (Figure 3b). Hydroxyl groups in the tetracyclic chain still play a key role in binding EptA residues where they form a hydrogen bond with TPO280 and salt bridge with ZN2. Hydrogens of amine groups in the tail contribute to establishing hydrogen bonds with Asp452 and His453. Other hydrogen bonds are also observed between Ligand11 with several EptA residues such as Val238, Thr242, Lys328, and Ser451. Interestingly, this ligand also forms pi-cation interactions between two positively-charged nitrogen atoms of different amine groups with the indole ring of Trp320 and the phenol ring of Tyr449, respectively.

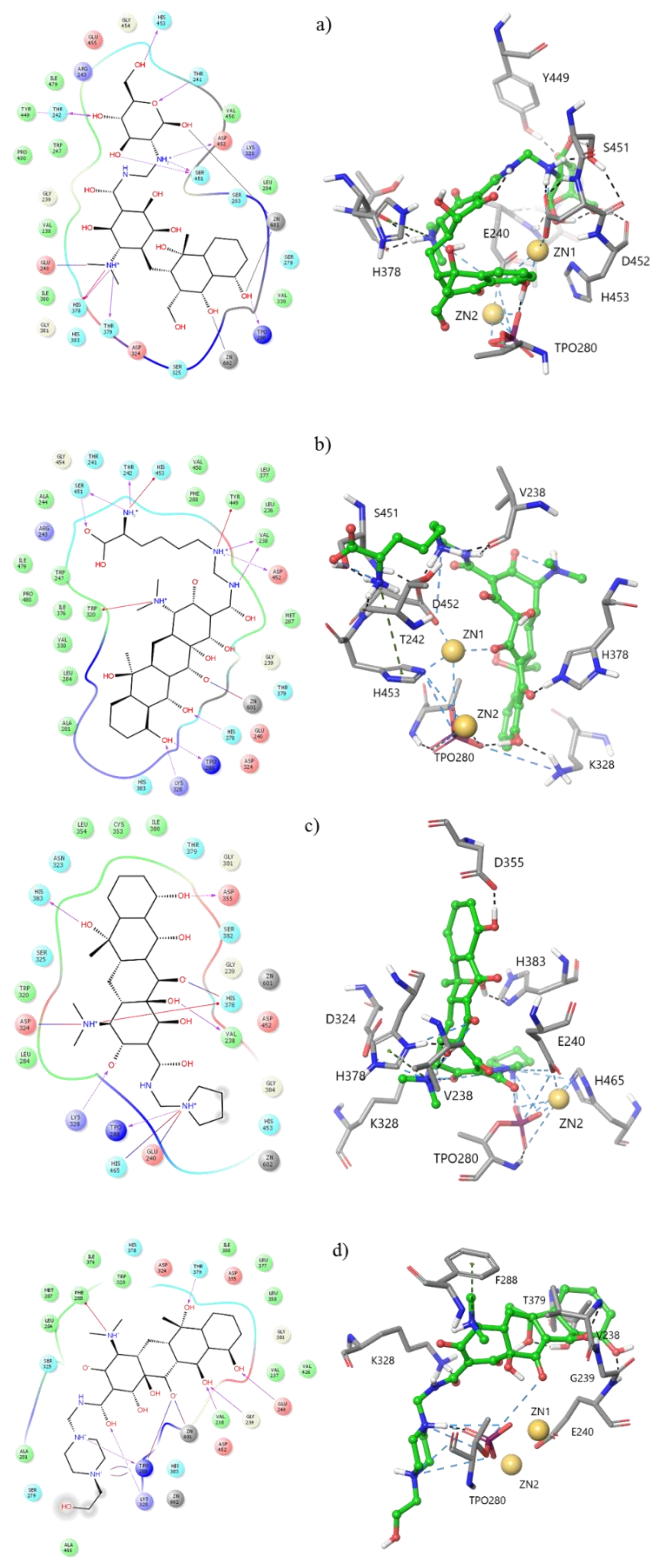

**Figure (3)**. Two-dimensional (left) and three-dimensional (right) interaction analyses of: a) Ligand4; b) Ligand11; c) Ligand17; and d) Ligand20. Graphical legends are the same as in Figure (2).

Distinct interaction poses are observed in the other two ligands, where the tetracycles face outward of catalytic cavity rather than inward as in Ligand4 and Ligand11. This position causes the last two ligands, Ligand17 and Ligand20, unable to bind zinc ions due to lack of negatively-charged atoms in their tail to form electrostatic interactions with metal ions. The inability to form salt bridges with zinc ions significantly reduced these two ligands Gibbs free energy of binding.

Ligand17 is registered as rolitetracycline and has a pyrrolidine ring attached to its main tetracyclic ring. Although unable to form an electrostatic contact with either zinc ions, the positively-charged pyrrolidine nitrogen of the ligand establishes a salt bridge with the catalytic Glu240 (Figure 3C). The polar hydrogen in the pyrrolidine ring also forms a hydrogen bond with the oxygen of TPO280. Ligand17 fails to make interactions with other active residues Asp452 and His453. Nevertheless, hydroxyl groups in its tetracyclic chain maintain hydrogen bonds with several residues such as Val238, Lys328, Asp355, and His383. Ligand17 also creates another three electrostatic contacts with Asp324, His378, and His465. A pi-cation interaction is also observed between the positively-charged tertiary amine nitrogen with the imidazole ring of His378.

Similar to Ligand17, Ligand20 also unable to form an interaction with Asp452 and His453 as its tetracyclic rings facing outward the catalytic site (Figure 3D). Ligand20 is another tetracycline-derived drug commonly called pipacycline. The distinct feature of this antibiotic is its piperazine ring attached to the main ring of tetracycline backbone. The positively-charged nitrogen and the polar hydrogens in its piperazine ring provide both electrostatic interactions and hydrogen bonds with TPO280, respectively. On the one hand, unlike Ligand17, Ligand20 maintains an interaction with one of zincs, ZN1. This interaction is made by one of the polar oxygens of hydroxyl groups and the same oxygen atom also creates an electrostatic ring with the partially-positive phosphor atom of TPO280. Ligand20 also successfully manages to make a hydrogen bond with Glu240. Other interactions observed in the best pose of Ligand20 are hydrogen bonds with Val238, Gly239, Lys328, and Thr379. A pi-cation contact is also established between the cationic tertiary nitrogen with the phenolic ring of Phe288.

Overall, ligands derived from similarity search performed better in molecular docking assays and interaction analyses than the EptA natural substrate PEA. These tetracycline derived antibiotics have a lower free energy of binding, which represents the high spontaneity and stability of the protein-ligand complex formation. This result possibly due to a higher number of interactions made by the assessed ligands with the protein. Although some ligands are incapable to form interactions with several catalytically important residues such as Asp452 and His453, other strong interactions like hydrogen bonds, electrostatic bridges, and pi-cation interactions with the protein residues account for the high stability of the formed complex. The ligands also have more hydrophobic contacts with the protein than PEA. Although this type of interaction has a much lower energy than the aforementioned interactions, a higher number of hydrophobic contacts could significantly contribute to overall ligand binding to the protein receptor.

### 3.4.2. Interaction analyses of pharmacophore-based compounds

Compounds resulted from ZincPharmer pharmacophore screening and validated by molecular docking have a various binding mode with the receptor. This varied binding mode is influenced by the type of molecules, specifically chemical groups attached in their structure. Ligand9 and Ligand15 are similar in structure in which they consist of three chained sugar or glycoside molecules enriched with hydroxyl and amine groups. A further search of Ligand9 and Ligand15 in ZINC database revealed that these two are well-known antibiotics. Ligand9 is gentamicin and Ligand15 is kanamycin which both belong to aminoglycoside antibacterials. Aminoglycosides are well-known antibiotics to treat wide spectrum bacterial infection [51].

The docking with EptA demonstrates that Ligand9 and Ligand15 have a comparable interaction pose (Figure 4B and 4C). They create a number hydrophobic contacts as well as strong interactions with catalytically active residues such as TPO280 and Glu240. Ligand9 creates a hydrogen bond between the polar hydrogen of a hydroxyl group with the negatively-charged oxygen of TPO280 phosphate group. A hydrogen bond is also established between the polar hydrogen of an amine group with the carbonyl oxygen of Glu240. A similar pattern of interaction is also shown by the Ligand15. Two of its polar hydrogens makes hydrogen bonds with oxygens of TPO280. Whereas Glu240 primary chain of EptA serves as both donor and acceptor of hydrogen bonds with Ligand15. However, these ligands are unable to make interactions with both zinc cofactors and the other two catalytic residues Asp452 and His453. This may result in an unfavorable inhibition mode against EptA which also explain their low binding score.

Significantly more extensive interactions with the receptor are shown by Ligand3 and Ligand19. Unlike previously described pharmacophore-derived compounds which are identified as antibiotics, Ligand3 and Ligand19 belong to nucleoside compounds: guanosine diphosphate (GDP) and deoxyguanosine diphosphate (dGDP), respectively. The distinguishing factor of these ligands to the other is the presence of phosphate groups which mimic the structure of PEA as the natural substrate of EptA. Phosphate groups possessed by these ligands greatly affect the binding with EptA by creating strong interactions with all catalytic residues and zinc ions.

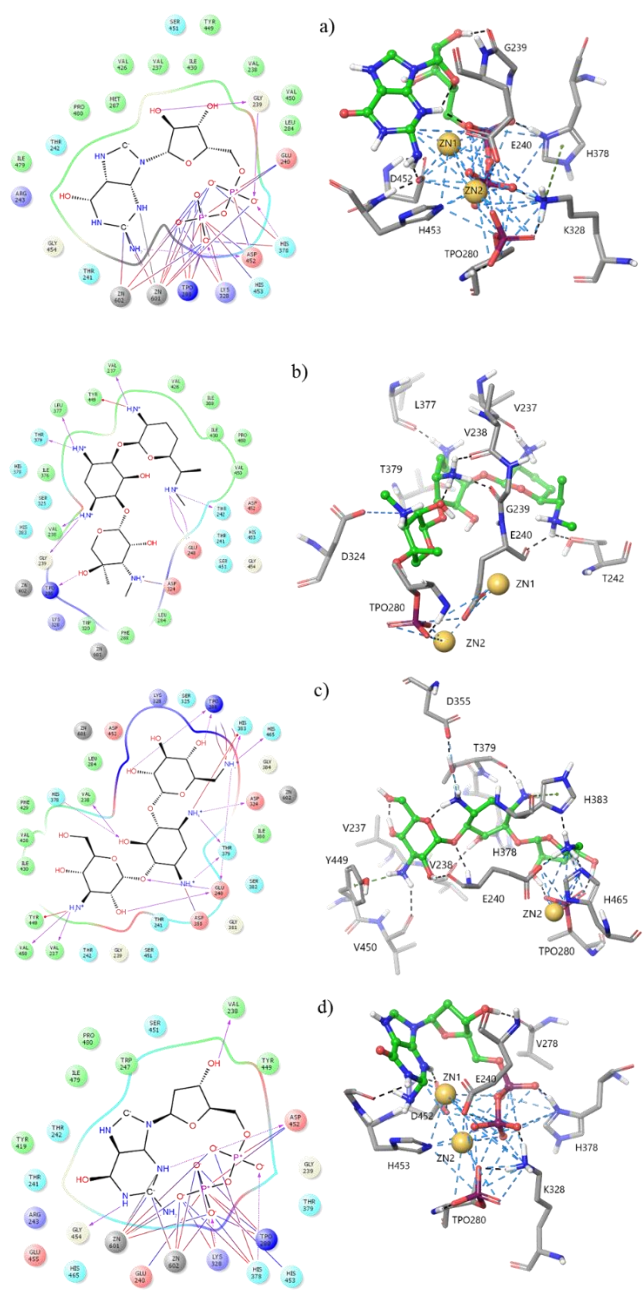

**Figure (4)**. Two-dimensional (left) and three-dimensional (right) interaction analyses of: a) Ligand3; b) Ligand9; c) Ligand15; and d) Ligand19. Graphical legends are the same as in Figure (2).

In Ligand3 for instance, negatively-charged oxygens of the phosphate groups create a salt bridge with both $Zn^{2+}$ ions and the partially positive phosphor atom of TPO280. In addition, two positively-charged phosphors of both phosphates are able to make electrostatic bridges with all catalytic triad residues. The same interaction pattern is also apparent in the Ligand19 where its oxygens and phosphors of phosphate groups contribute to the interaction with all catalytic triad and zinc cofactors. Both ligands also have bent position which allows their hydrophobic moieties to form hydrophobic contacts with the protein catalytic cavity wall. These interactions create a binding network that could possibly increase the binding affinity of ligands, decrease their free energy of binding with the protein receptor, and increase the overall stability of complex formed by the target protein and ligands.

### 3.5. Combined pharmacophore and ligand-guided virtual screening for drug discovery

A combination of pharmacophore and substrate similarity-based virtual screening was employed in this study to search potential EptA inhibitors. Pharmacophore screening has been an established approach to virtual screening in drug discovery and medicinal chemistry in general [52]. This technique utilises an algorithmic search of specific spatial and chemical arrangement of a compound or pharmacophore in the compound database. In the present study, pharmacophore screening was performed in ZincPharmer [34] platform (http://zincpharmer.csb.pitt.edu) which uses the former Pharmer [46] algorithm to screen the pharmacophore of compounds in ZINC database [36]. ZincPharmer screening can be employed to search matched compounds using crystal structures of a protein in complex with ligands that are available in PDB. However, as the crystallographic structure of EptA bound with its natural substrate PEA is not available anywhere in protein structure databases, the re-docked EptA-PEA complex structure was used in the screening. In addition to the pharmacophore screening, PEA structural ligand-based similarity search was also done to enhance breadth and diversity of assessed ligands as well as to compare and combine both approaches. The search was done in PubChem [37] by using its structure search tool (available in https://pubchem.ncbi.nlm.nih.gov/). Some antibiotics and antibacterials were also included in the library. This combined screening produced a library of 8166 hits for further molecular docking validation.

Molecular docking validation was done in two independent step using different scoring and refinement parameters. This approach was applied as a duplication to refine docking result and to reduce bias. Best 20 ligands were selected based on their binding energy upon docking with the EptA catalytic site, which are lower than the EptA natural substrate PEA. Interaction analyses of selected ligands demonstrated that most ligands maintain essential interactions with EptA. These ligands have strong interactions with at least two catalytically important residues whilst creating other interactions with some EptA residues which contribute to their low binding free energy with EptA.

A structural search revealed that these ligands are well-known antibiotics or active molecules such as nucleosides. Four selected ligands from PubChem database: Ligand 4, 11, 17, and 20 belong to a group of antibiotics called tetracyclines, named meglucycline, lymecycline, rolitetracycline, and pipacycline, respectively. Tetracyclines have four linearly adjacent carbon rings as their backbone structure and are well-established antimicrobes used to treat a wide range of microbial infections and diseases [50]. The antibacterial activity of tetracyclines is mainly due to their action in preventing the attachment of aminoacyl-tRNA to the ribosomal acceptor site during mRNA translation and protein synthesis [49,50]. Although the protein synthesis inhibition activity of tetracycline is well studied, their activity against other potential targets is still inadequately studied.

The present study demonstrates that tetracyclines can potentially be re-purposed to combat antibiotic resistance in Gram-negative bacteria by inhibiting one of the resistance factors lipid A–PEA transferase (EptA). Tetracycline ligands have lower binding free energy and overall better binding capacity to the EptA catalytic site than PEA. Hydroxyl groups in their tetracyclic ring may electrostatically bind zinc ions that are essential for EptA catalytic activity. This finding is in accordance with earlier experimental studies showing that tetracyclines have a strong metal-chelating activity that affects their antimicrobial activities [53,54].

ZincPharmer pharmacophore search returned more varied hit ligands. Two of selected ligands (Ligand 9 and 15) are known antibiotics gentamicin and kanamycin, respectively. These two compounds belong to the same group of antibiotics called aminoglycosides which are composed of three sugar molecules or glycosides that have amine group modifications. Even though these ligands enriched in polar groups such as hydroxyls and amines, they failed to make electrostatic contacts with zinc cofactors. This is probably due to their orientation during molecular docking where negative charge bearing atoms such as oxygen and nitrogen could not reach a sufficient distance and a suitable position to establish contacts with zinc ions.

On the other hand, other selected ligands from ZincPharmer screening, Ligand 3 and 19 have notable interactions with EptA catalytic site. Ligand 3 and 9, known respectively as guanosine diphosphate (GDP) and deoxyguanosine diphosphate (dGDP) are examples of bioactive molecules belong to nucleosides. These two compounds successfully create multiple interactions with all EptA catalytically active residues, including the catalytic tetrad: Glu240, TPO280, Asp452, and His453 and two zinc cofactors: ZN1 and ZN2. These binding networks are mainly created by the presence of phosphate groups in their structure, enabling them to mimic the EptA natural substrate PEA and bind all pivotal EptA residues. This result demonstrates that molecules derived from pharmacophore screening can match the spatial and chemical arrangement of the template molecule and even enhance their pharmacological characteristics such as protein binding capacity.

These results show that some known antibiotics could possibly be re-purposed to combat Gram-negative bacteria infection by inhibiting their resistance factor EptA. Essential modifications, such as chemical group replacement, of these antibiotics might be necessary to enhance their susceptibility to bind and inhibit EptA catalytic activities. Furthermore, interaction analyses of two nucleoside ligands (Ligand 3 and 19) show that these compounds create multiple interactions with all EptA active residues. These ligands can be further tested for their inhibition activity against EptA and be developed as new potent inhibitors against the antibiotic resistance of Gram-negative bacteria.

## CONCLUSION

Gram-negative bacteria infection remains a major global health concern due to a rapid development of multi-drug resistant bacterial strains. The structural characterisation of a Gram-negative bacteria resistance factor lipid A–PEA transferase (EptA) of *Neisseria meningitidis* has opened a new possibility to combat the resistance by developing potent compounds that inhibit its activity. In this study, a combined pharmacophore and ligand-guided high throughput virtual screening were performed to screen potential inhibitors from millions of commercially available and registered compounds in databases.

Initially, more than 8,000 compounds derived from both pharmacophore screening in ZincPharmer and ligand-based similarity search in Pubmed subjected to molecular docking validation and refinement. The validation resulted in best 20 consensus ligands that have a lower free energy of binding than EptA natural substrate PEA. Further interaction analyses suggest that eight selected ligands (Ligand 3, 4, 9, 11, 15, 17, 19, and 20) have overall more interactions with the receptor than PEA and maintain essential interactions with catalytically active residues and cofactors as well as contacts with other residues. Some ligands notably have distinct binding mode than others, i.e. Ligand3 and Ligand19 that capable of creating a binding network with all EptA catalytic tetrad and zinc cofactors. Pharmacophore and ligand similarity-based compounds from this study can be further studied to evaluate their inhibition capability *in vitro* and *in vivo* and further developed into novel antibacterial agents against multi-drug resistant Gram-negative bacteria.

## CONFLICT OF INTEREST

The author declares that there is no conflict of interest regarding the publication of this paper.

## ACKNOWLEDGEMENTS

The author gratefully acknowledges the Indonesian Endowment Fund for Education (LPDP) the Republic of Indonesia for their scholarship through Indonesian Educational Scholarship (BPI) scheme and their support towards the publication of this study.